# Growth of epitaxial ReS$_2$ thin film by pulsed laser deposition

B. Vishal,[1,*] H. Sharona,[1] U. Bhat,[1] A. Paul,[1] M. B. Sreedhara,[1] V. Rajaji,[1,3] S. C. Sarma,[2,3] C. Narayana,[1,3] S. C. Peter,[2,3] and R. Datta[1,*]

[1]*International Centre for Materials Science, Chemistry and Physics of Materials Unit, Jawaharlal Nehru Centre for Advanced Scientific Research, Bangalore 560064, India.*

[2]*New Chemistry Unit, Jawaharlal Nehru Centre for Advanced Scientific Research, Bangalore 560064, India.*

[3] *School of Advanced Materials, Jawaharlal Nehru Centre for Advanced Scientific Research, Bangalore 560064, India.*

**Abstract**

We present results on growth of large area epitaxial ReS$_2$ thin film both on *c* plane sapphire substrate and MoS$_2$ template by pulsed laser deposition (PLD). Films tend to grow with $(0001)ReS_2 \perp (0001)Al_2O_3$ and $(0001)ReS_2 \perp (0001)MoS_2 \parallel (0001)Al_2O_3$ at deposition temperature below 300°C. Films are polycrystalline grown at temperature above 300°C. The smoothness and quality of the films are significantly improved when grown on MoS$_2$ template compared to sapphire substrate. The results show that PLD is suitable to grow ReS$_2$ epitaxial thin film over large area for practical device application.

[*]Corresponding author e-mail: *ranjan@jncasr.ac.in, badriv@jncasr.ac.in*

# I. Introduction

ReS$_2$ belongs to the class of material known as two dimensional transition metal dichalcogenides (2D TMDs) [1-5]. The monolayer structure of ReS$_2$ is Peierls distorted 1T$_d$ ($space\ group: P\bar{1}$) where four Re atoms are hybridized together to form 2D periodic superstructure [Figure 1]. ReS$_2$ has unique characteristic compared to other TMDs in the family e.g., MoS$_2$, WS$_2$ etc.; it does not undergo direct to indirect band gap transition in more than mono layer form. This has been attributed to weak interlayer van der Walls coupling i.e. ~18 meV compared to 460 meV in case of MoS$_2$ [5]. This is beneficial in many applications where retention of optical direct type band gap is desired in two or few layer form of such materials. The band gap of ReS$_2$ is 1.52 and 1.42 eV in monolayer and multilayer form, respectively [6]. ReS$_2$ have potential applications in electronic, optoelectronic device fabrication, energy storage, and catalytic hydrogen production. Ultra-thin 2D materials provide a solution to overcome the issues associated with continuous demand on scaling-down of transistors. ReS$_2$ based thin film transistors (TFT), field effect transistors (FET) digital logic devices, and photodetectors have been demonstrated [7-9]. Due to its inherent modulated structure, ReS$_2$ exhibits unique anisotropic optical and electrical properties compared to most other 2D materials in the family. Due to its optical anisotropy property, ReS$_2$ is already in use as polarization dependent photodetectors [4, 6, 8]. Anisotropy dependent photo current has been demonstrated in ReS$_2$ which shows promise in integrating electronic and optical devices e.g., as phototransistor part of an integrated circuit [9]. Maximum anisotropic ratio of 3.1 is reported in ReS$_2$ based FET devices with competitive values of current on-off ratio and low subthreshold swing [7, 10]. ReS$_2$ can be used as efficient photocatalyst. Ab initio calculations shows that both mono and multilayer ReS$_2$ are stable and efficient photocatalyst for water splitting [11]. ReS$_2$ on carbon is found to be promising electrocatalyst for hydrogen evolution reaction (HER) and suggested that perpendicular orientation of ReS$_2$ flake will improve the performance further [12-13].

Therefore, to exploit the unique properties of ReS$_2$ development of its synthesis procedures in the form of large area mono and few layers is required. Monolayer and few layers of ReS$_2$ can be synthesized by both bottom up methods: chemical vapor deposition (CVD), atomic layer deposition (ALD), pulsed laser deposition (PLD), physical vapor deposition (PVD) and top-down methods: mechanical, chemical and liquid exfoliations techniques [14-17]. However, the exfoliation methods come with the challenge of controlling the dimension of flakes, contamination or residues of exfoliation agent and misalignment in stacking of two or more similar or different TMDs while fabricating heterostructure [4, 18]. For device application, it is important to deposit 2D materials and their heterostructures by crystal growth techniques over large area uniformly [2, 19, 20]. There are already reports on such attempts to grow large area ReS$_2$ thin film by physical and chemical deposition techniques. In this context, ReS$_2$ was grown by CVD on Si/SiO$_2$ substrate at 450°C with a mean crystal size of 40 µm. Ammonium per-rhenate and sulphur were used as precursors and argon as carrier gas. Depending upon the temperature of the growth different size and thickness of the crystal are obtained. The nucleation of the secondary layer initiated at the centre of the primary crystal and growth proceeded towards the edge [15]. Te assisted ReS$_2$ growth was performed in a CVD reactor at 700°C. Te helps in lowering the melting point ( 3180 °C) of Re thus enhances its incorporation, which is not the problem with other components having relatively lower melting points [16]. ReS$_2$ was grown by CVD on a flexible glass substrate at 450 ºC in the absence of catalyst [21]. The ReO$_3$ powder was used as a precursor which was sulphurized by H$_2$S for 30 min. Large area uniform ReS$_2$ thin film was grown by CVD on mica substrate at 600ºC under ambient pressure. Re metal powder and Re$_2$S$_7$ were used as precursors. Re$_2$S$_7$ is volatilized above 300ºC, and partial sulphurization provides nucleation sites on the substrate for the subsequent growth of ReS$_2$ Chemically inert mica substrate reduces nucleation density, atom migration and prevent vertical growth of ReS$_2$ compared to SiO$_2$ substrate where growth is predominantly vertical [22, 23]. The film grown on mica substrate grows with optimum flat morphology below 700 ºC and start forming amorphous domains at temperature below 500 ºC. Few layers centimetre scale continuous ReS$_2$ film was grown on SiO$_2$/Si substrate by physical vapour deposition using ReS$_2$ powder as source material [24]. The ReS$_2$ film was grown by ALD over large area (5 sq. cm) on Al$_2$O$_3$ coated Si (100) substrate at wide temperature range between 120-500°C. The morphology of films was rough containing both horizontal (*c* plane) and tilted vertical domains. The perpendicular orientation of ReS$_2$ was found to be beneficial for (hydrogen evolution reaction) HER and Li-ion and Li-S batteries though for device application flat orientation is required [13, 25-27].

As already pointed out that from the practical applications point of view, it is desirable to grow/deposit such 2D materials uniformly over large area on a substrate with a simpler technique involving less chemicals and process parameters. In this context, large area epitaxial growth of MoS$_2$ and WS$_2$ thin film with control over layer numbers was demonstrated by pulsed laser deposition [19]. In the present report, the growth of large area ReS$_2$ thin film is investigated by PLD. The major difference between ReS$_2$ and earlier growth of MoS$_2$ (and WS$_2$) by PLD is the melting point of the material: 390 ºC compare to 1185°C (1250°C) of MoS$_2$ (WS$_2$). Two different substrates were used in the present investigation: *c* plane sapphire substrate and MoS$_2$ template grown on *c* plane sapphire. It is found that depending on the growth temperature the film can be vertically oriented, a composite of vertical and ploycrystalline, and completely polycrystalline films. The quality of the layer improves significantly while grown on MoS$_2$ template. We present detailed structural and phase characterizations and optical property evaluation by HRTEM, X-ray, Raman, and CL techniques.

## II. Experimental Methods

ReS$_2$ thin films were grown by PLD on two different substrates; $c$ plane sapphire substrate and 20 nm thick $c$ plane MoS$_2$ template grown on $c$ plane sapphire substrate following the procedure already described in Ref. 19. ReS$_2$ target pallet was used to ablate and grow the thin film. ReS$_2$ compound was first synthesized by directly heating mixture of pieces of elemental Re cut from the wire (1mm diameter, 99.97%, Alfa Aesar) and S powder (-325 mesh, 99.5% Alfa Aesar) in the stoichiometry ration (1:2) sealed in a quartz tube at $10^5$ mbar pressure to 200 ºC in 15 h followed by annealing at that temperature for 2 h. This step involving slow heating necessary to avoid possible explosion due to the high vapour pressure of S. Then the temperature was increased stepwise to 400 °C and finally to 900 °C over a period of 10 h and 24 h, respectively. The sample was held at 900 ºC for 120 h to ensure homogeneity. The black powder thus obtained was then cold pressed in a die and vacuum sintered at temperature 200 °C for 4 h to make PLD target. The temperature and sintering duration were comparatively lower compare to MoS$_2$ and WS$_2$ pellets due to low melting temperature (380°C) of ReS$_2$. The sintering in a vacuum chamber with continuously running pump in the background prevents oxidation of compound and deposition of vapour specifics back on the pallet surface compared to the case for sintering performed in a sealed quartz tube. The deposition of the entire ReS$_2$ thin film were carried out at the same temperature and pressure ($10^{-5}$ Torr) with laser ablation frequency of 1 Hz. Slow laser ablation rate allows sufficient time for kinetic relaxation of the nucleation layer which helps in establishing the epitaxial relationship with the substrate thus removing misaligned crystallite. This was already demonstrated in the case of ZnO and its alloys, MoS$_2$ and WS$_2$ [19]. The ReS$_2$ thin film was grown at various deposition temperatures: 100 to 400 °C and is summarized in the Table 1. The PLD method of growth has advantages over CVD and PVD counterparts in the sense that the films may be free from any parasitic deposition. The method is scalable and economical.

All the films were investigated by high resolution transmission electron microscopy (HRTEM), X-ray diffraction, Raman spectroscopy, and cathodoluminescence (CL). HRTEM images were recorded in a double aberration corrected FEI TITAN 80-300 kV microscope. TEM cross-sectional specimens were prepared by first mechanical thinning to 90 µm and then tripod polishing down to 20 µm. The final thinning to perforation was carried out in a Gatan PIPs (precision Ar ion polishing system, Gatan, Inc.) operating at 4.5 kV to generate electron transparent thin area. Special care was taken during the sample preparation to prevent delamination of weak van der Waals layers from the substrate. Raman spectra were recorded using a custom-built Raman spectrometer using a 532 nm laser excitation and 1800 lines/mm grating at room temperature [28] CL measurement was carried out using Gatan mono CL (serial mode) in a FEI quanta FE-SEM. All the CL spectra were collected with Peltier cooled (-25ºC) photomultiplier tube (PMT) with 1 nm step size and accusation time of 0.6 s/step. Entrance and exit slits size of the spectrometer was kept at 1 mm.

### III.     Results and Discussion

We start describing the structural characterization of $ReS_2$ thin films by transmission electron microscopy. Fig. 2 (a)-(f) are the low magnification cross sectional bright field TEM images of $ReS_2$ thin film on *c* plane sapphire at various growth temperatures: 150, 200, 250, 300 and 400°C respectively. The thickness of the films is in the range of ~ 30-50 nm except the film grown at 150 °C (300 nm). The films are smooth and continuous but at discrete places on the surface 3D growth of crystals can be observed turning the film rough. Formation of 3D crystals on the surface is more predominant at lower deposition temperatures and suppresses with increasing growth temperature till 300°C. Scattered dark particles can be observed throughout the volume of the film and are found to be Re metals. Next, details are provided on the crystal orientation of the thin film with respect to the substrate, identification of the dispersed dark particle and nature of surface 3D crystals by HRTEM.

Fig. 3 (a) is the HRTEM image of the ReS$_2$ thin film (150°C) along $<2\bar{1}\bar{1}0>$ zone axis (Z.A.) of sapphire. Both the image and corresponding FFT pattern confirm that the (0001) plane of ReS$_2$ is perpendicular to the (0001) plane of sapphire. The epitaxial relationship is thus $(0001)ReS_2 \perp (0001)Al_2O_3$. The vertical planes started forming at the interface and continue towards the surface. Scattered particles with dark contrast marked with white circle can be observed throughout the film. These particles are identified as hexagonal Re metals (space group P63/mmc, d$_{0002}$ = 2.22 Å) [Fig. 3 (b)]. The appearance of particles is dark because Re has higher average atomic number (Z = 75) per unit volume compared to ReS$_2$ thus scattering more electrons out of the aperture (mass-thickness contrast). The lattice parameters of hexagonal Re is $a$ = 2.76 Å and $c$ = 4.45 Å. The size of the particles are ~ 30 nm. The 3D crystals formation on the surface for film grown at 150 °C leading to rough morphology is also found to be hexagonal Re metals [Fig. S1]. The possible origin of Re metals could be from the ReS$_2$ target pellet. During vacuum sintering there is a prospect for S loss and formation of local Re rich areas which may subsequently form Re metals nano particles during laser ablation and dispersed throughout the film. XRD of ReS$_2$ pallet reveals the presence of Re metal in the pellet [Fig. S3]. Therefore, formation of Re metal particles in the film volume can be avoided if ReS$_2$ target pellets can be fabricated without such Re rich local regions and this may be done with an improved sintering process preventing S loss. In some places, the films are found to be amorphous coexisting with the crystalline regions [Fig. S4]. The film grown at 200 °C shows similar features as 150 °C grown film but from place to place the *c* planes start branching away from the vertical direction and creating a dome like structure at the surface [Fig. 3 (c) & Fig. S5]. In some areas, Re metal with cubic crystal structure is detected though the hexagonal form is the predominant phase [Fig. 3 (d)]. The lattice parameter of metastable fcc Re is 3.89 Å. The formation of metastable Re phase might have been due to phase transformation from hcp to fcc structure following Bain transformation [29]. With increasing deposition temperature to 250 °C, the branching start dominating near the interface [Fig. 3(e)]. This led to the formation of films consisting of both vertical domains and tilted domains. This is also reflected in the X-ray diffraction pattern in terms of additional peaks [Fig.5]. However, the most favoured vertical orientation is maintained, and domains of rotating vertical planes regions can be observed. The region near the interface of the film grown at 300 °C is polycrystalline made of mostly $ReS_2$ crystal (~ 5 nm size) with vertical planes growing on top of it meandering towards the surface [Fig. 3 (f) & (g)]. Finally, the film grown at 400 °C, which is above the melting point of $ReS_2$ is completely polycrystalline and made of Re metals [Fig. 3 (h)].

Fig. 4 (a)-(c) are low magnification TEM images of the $ReS_2$ thin film grown on $MoS_2$ template at three different temperatures: 150, 250, and 400 °C, respectively. The thickness and growth temperature of the $MoS_2$ template layer is 20 nm and 400 °C, respectively following the procedure already described in Ref. 19. Two different layers can be identified from the mass-thickness contrast. $ReS_2$ appears darker compared to $MoS_2$ due to higher molecular mass. The surface and interface are smooth and free from dome like morphology compared to films directly grown on the sapphire substrate except for the film grown at 400 °C. The thickness of the $ReS_2$ films is ~ 20 nm. Fig. 4 (d)-(f) are HRTEM images from three different $ReS_2$ films on $MoS_2$ template. From the image and the FFT pattern it can be observed that the *c* plane of $MoS_2$ film is parallel to the interface whereas in case of $ReS_2$ the *c* plane is perpendicular for films grown at 150 and 250 °C. The epitaxial relationship is the same as in the case of sapphire. The nucleation of $ReS_2$ occurs at the terminating edges of the $MoS_2$ planes and propagates along the vertical direction. The film grown at 400°C is polycrystalline as the temperature is above the melting point of $ReS_2$ and consisting of Re metals which is similar to the case of film grown on sapphire substrate.

Fig. 5 is the X-ray diffraction pattern of $ReS_2$ films grown on both sapphire and $MoS_2$ template at various temperatures. The number of peaks is restricted due to specific orientation of both film and substrate. Sapphire (0006) peak appears at $2\theta = 41.8°$. The peak at 32.2° is from $(20\bar{2}0)$ plane of vertically oriented $ReS_2$ film. Other peaks of $ReS_2$ identified at 20.4, 27.3, and 38.1° corresponding to $(1\bar{1}02)$, $(11\bar{2}1)$, and $(02\bar{2}2)$, planes respectively. Hexagonal Re $(1\bar{1}01)$ peak found at $2\theta = 42.9°$ is dominating for film grown at 150 °C due to large thickness compared to other films. The peak at 40.4° marked with * could be from shifted (0002) plane of $ReS_2$ as many inclined orientations can be observed from the HRTEM images which could contribute intensity to this peak [Fig. S6].

Raman spectra of all the $ReS_2$ films grown both on top of sapphire substrate and $MoS_2$ template are given in Fig. 6 (a). The modes corresponding to $ReS_2$ and $MoS_2$ are indicated in the figure. Thirteen Raman actives modes of $ReS_2$ are observed. These are five in-plane ($E_g$), four out-of-plane ($A_g$) and in-plane and out-of-plane coupled ($C_p$) vibrational modes. No modes corresponding to $ReS_2$ are observed for the film grown at 400 °C. This is consistent with the observation of Re metals in the films grown at 400 °C. The $E_g$ modes at 150.5, 161.3, 212.4 and 233.2 $cm^{-1}$ are due to in-plane vibrations of Re atoms, while the modes at 304.8 $cm^{-1}$ is due to in-plane vibrations of S atoms. $A_g$ modes located at 138.5 and 141.7 $cm^{-1}$ are due to out-of-plane vibrations of Re atoms, while the modes at 418.7 and 437.4 $cm^{-1}$ are from the out-of-plane vibrations of S atoms. The $C_p$ modes at 278.3 $cm^{-1}$ are due to couple in and out-of-plane vibration of Re and S atoms, while 320.6, 324.9 and 407.3 $cm^{-1}$ are the couple in-plane and out-of-plane vibration of S atoms [4, 30]. In case of $MoS_2$ characteristic $A_{1g}$ and $E^1_{2g}$ modes located at 378.8 and 405.8 $cm^{-1}$ are indicated. $E^1_{2g}$ modes located at 378.8 $cm^{-1}$ is due to the vibration of two S atoms in the opposite direction, while $A_{1g}$ mode located at 405.8 $cm^{-1}$ is due to the out-of-plane vibration of S atoms in opposite directions with respect to the Mo atom [31].

Fig. 6 (b) is the CL spectra of ReS$_2$ grown on sapphire and MoS$_2$ template. The band edge emission is observed to vary between 1.49 to 1.46 eV for the film grown directly on sapphire depending on the growth temperature. The films grown on MoS$_2$ template show emission peak around 1.46 eV. CL confirms that all the thin films are optically active material except for the films grown at 400 °C which consists of Re metals and may find wide applications. However, the challenge remains on how to grow a flat ReS$_2$ film by PLD, may be a choice of substrate or control over kinetic or process parameters need to be explored further.

## IV. Conclusions

The growth of ReS$_2$ thin films on sapphire substrate and MoS$_2$ template by PLD has been investigated at various deposition temperatures. The films grow *c* plane perpendicularly with respect to the film substrate interface on both the substrate at lower deposition temperatures. The films are polycrystalline and contain mostly Re metals grown at temperature above 400 °C. The quality of the film is better grown on MoS$_2$ template compared to sapphire substrate. The results demonstrate that PLD can be used to grow ReS$_2$ thin film for any possible device application.

**Acknowledgement**

The authors sincerely acknowledge ICMS and JNCASR for the research funding and advanced characterization facilities.

**Table 1.** List of samples deposited on two different substrates and various deposition temperatures. Nature of the films observed also mentioned.

| Substrates | Deposition temperature (°C) | Nature of Film |
|---|---|---|
| $c$ Al$_2$O3 | 150 | $(0001)ReS_2 \perp (0001)Al_2O_3$ |
| $c$ Al$_2$O3 | 200 | $(0001)ReS_2 \perp (0001)Al_2O_3$ with branching |
| $c$ Al$_2$O3 | 250 | $(0001)ReS_2 \perp (0001)Al_2O_3$ onset of polycrystal formation near interface |
| $c$ Al$_2$O3 | 300 | polycrystal layer near interface |
| $c$ Al$_2$O3 | 400 | Re metal polycrystalline film |
| $c$ Al$_2$O3/MoS$_2$ | 150 | $(0001)ReS_2 \perp (0001)MoS_2$ |
| $c$ Al$_2$O3/MoS$_2$ | 250 | $(0001)ReS_2 \perp (0001)MoS_2$ |
| $c$ Al$_2$O3/MoS$_2$ | 400 | Re metal polycrystalline film |

**Figure Captions:**

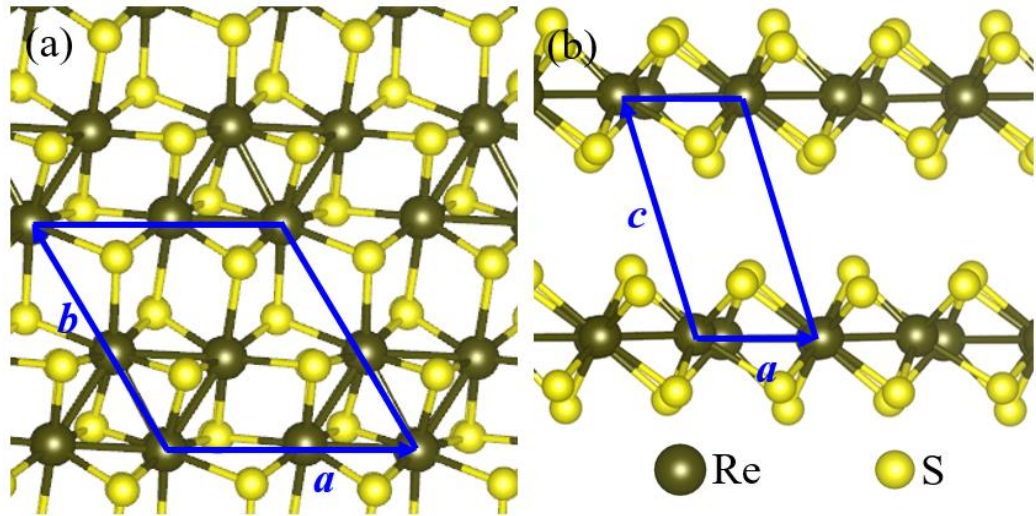

**Fig. 1**. Schematic structure of $1T_d$ ReS$_2$ (a) along <0001> Z.A. and (b) along <11$\bar{2}$0> Z. A. showing Bernal stacking [6]. The lattice parameters are: $a$ = 6.41 Å, $b$ = 6.47 Å, $c$ = 6.42 Å, α = 91.32°, β = 105.49° and γ = 119.03°.

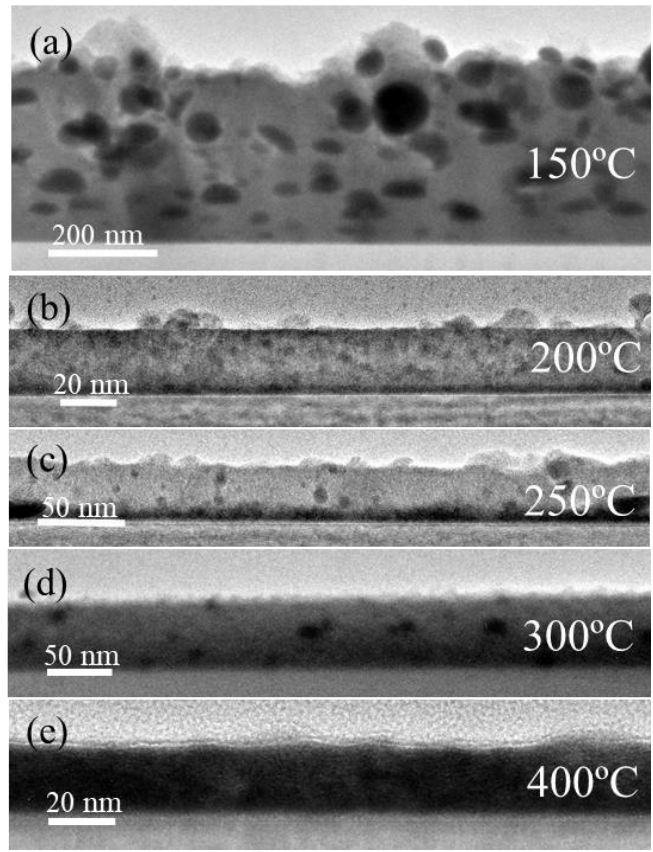

**Fig. 2**. (a)-(e) Low magnification cross sectional TEM bright field images of ReS$_2$ thin film grown on sapphire substrate at various deposition temperatures. The films are smooth and from place to place formation of 3D particles at the surface making the film rough. The thickness of the films is ~ 20-30 nm except the film grown at 150 °C (300 nm).

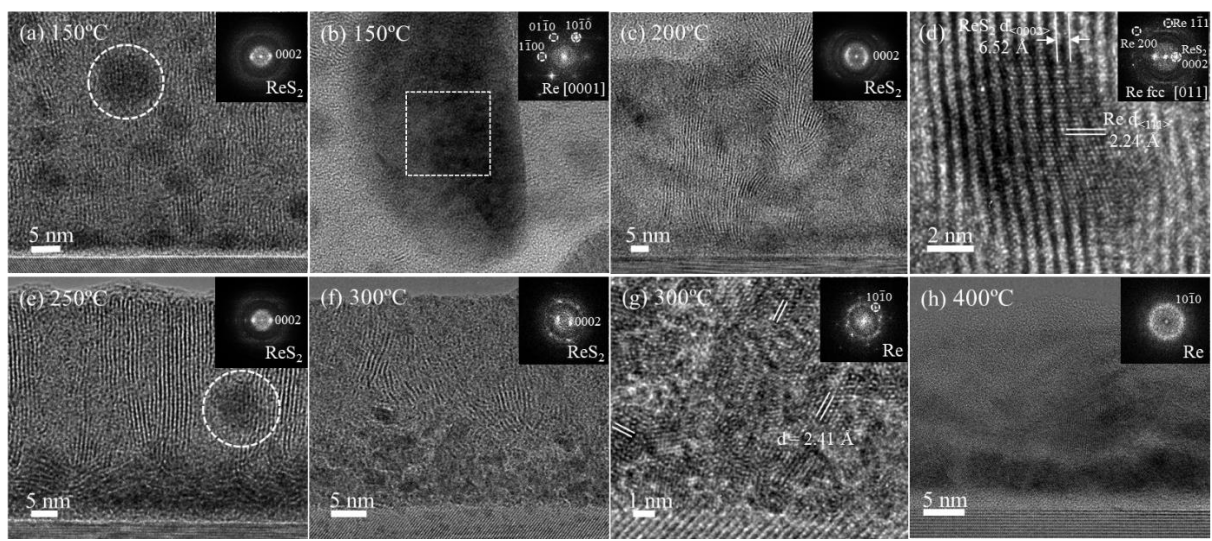

**Fig. 3.** HRTEM images of ReS$_2$ films grown on sapphire at various deposition temperatures. The films are grown with *c* plane perpendicular to the film substrate interface at (a) T = 150 °C, (b) Dispersed hexagonal Re metal particles are marked with dotted white squire. (c) & (e) Branching of ReS$_2$ layers can be seen in films grown at 200 and 250°C. (d) Metastable fcc Re metal (*a* = 3.89 Å) of ~3 nm size is observed. (f) & (g) film is polycrystalline grown at 250 °C near the interface and vertical planes above it is meandering towards the surface. (h) polycrystalline Re metal film is obtained at growth temperature of 400 °C.

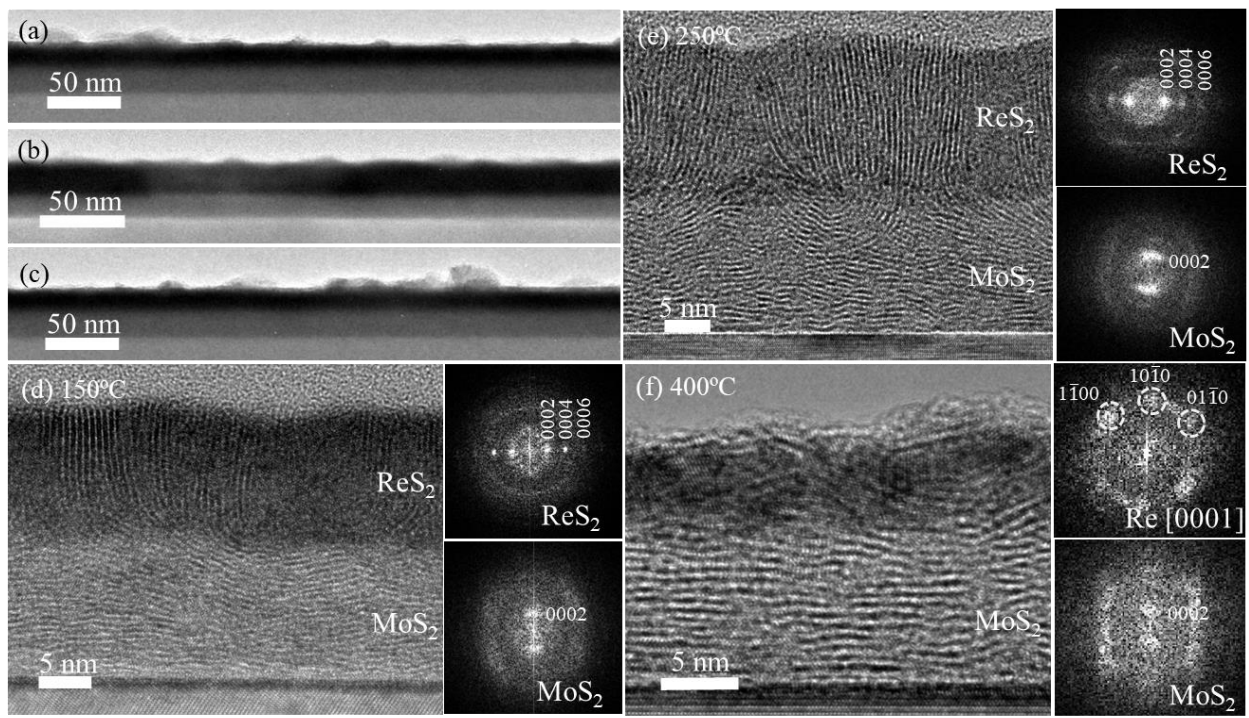

**Fig. 4.** (a)-(c) low magnification cross sectional TEM bright field images of ReS$_2$ films grown on MoS$_2$ template at three different temperatures: 150, 250 and 400 °C. Films are smooth for films grown at 150 and 250 °C. Roughness is more for film grown at 400°C. (e)-(f) HRTEM images from three different films. Epitaxial relationship is $(0001)ReS_2 \perp (0001)MoS_2 \parallel (0001)Al_2O_3$ for films grown at 250 and 300 °C. The film is polycrystalline at 400°C.

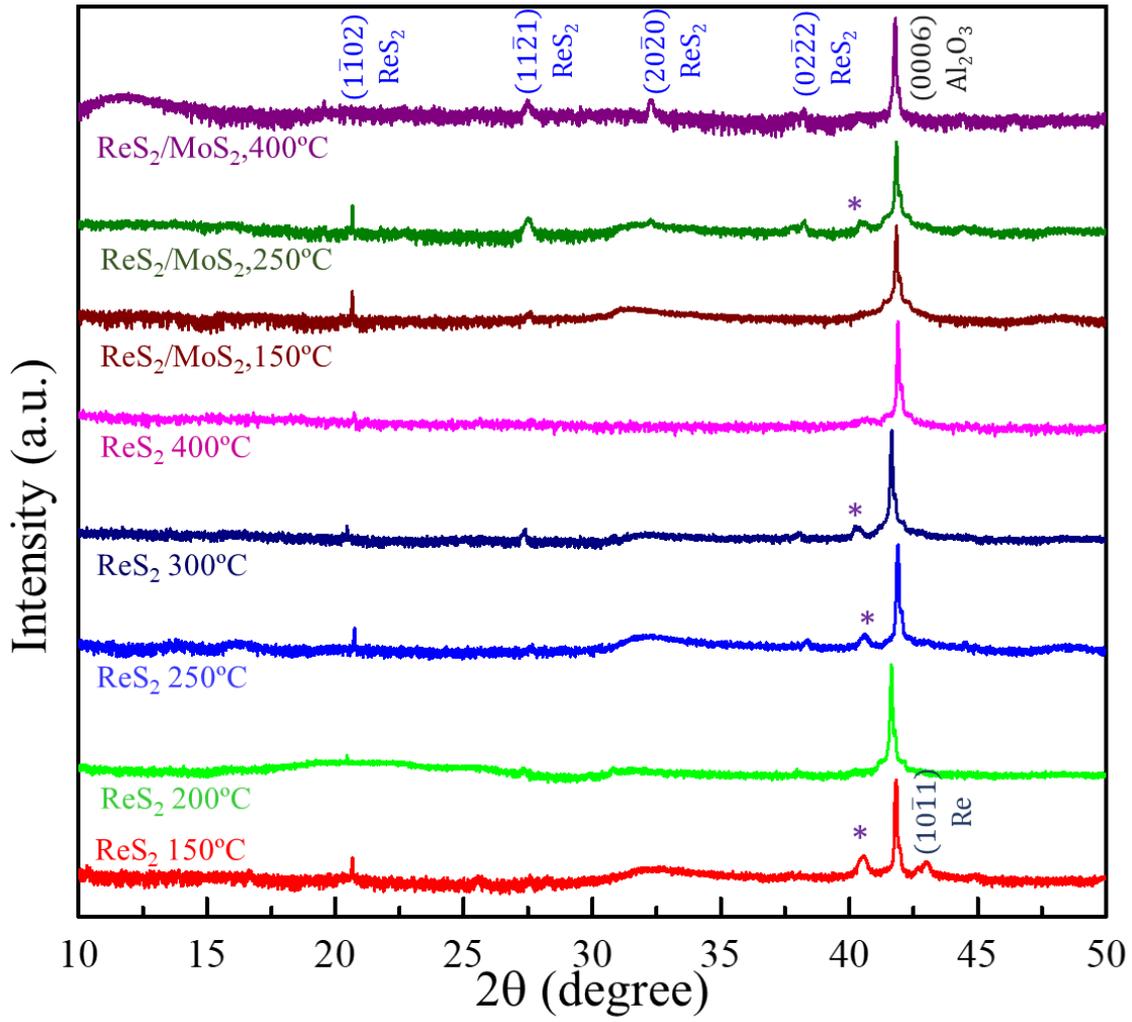

**Fig. 5.** X-ray diffraction pattern for various ReS$_2$ thin films grown on both sapphire substrate and MoS$_2$ template at different temperatures.

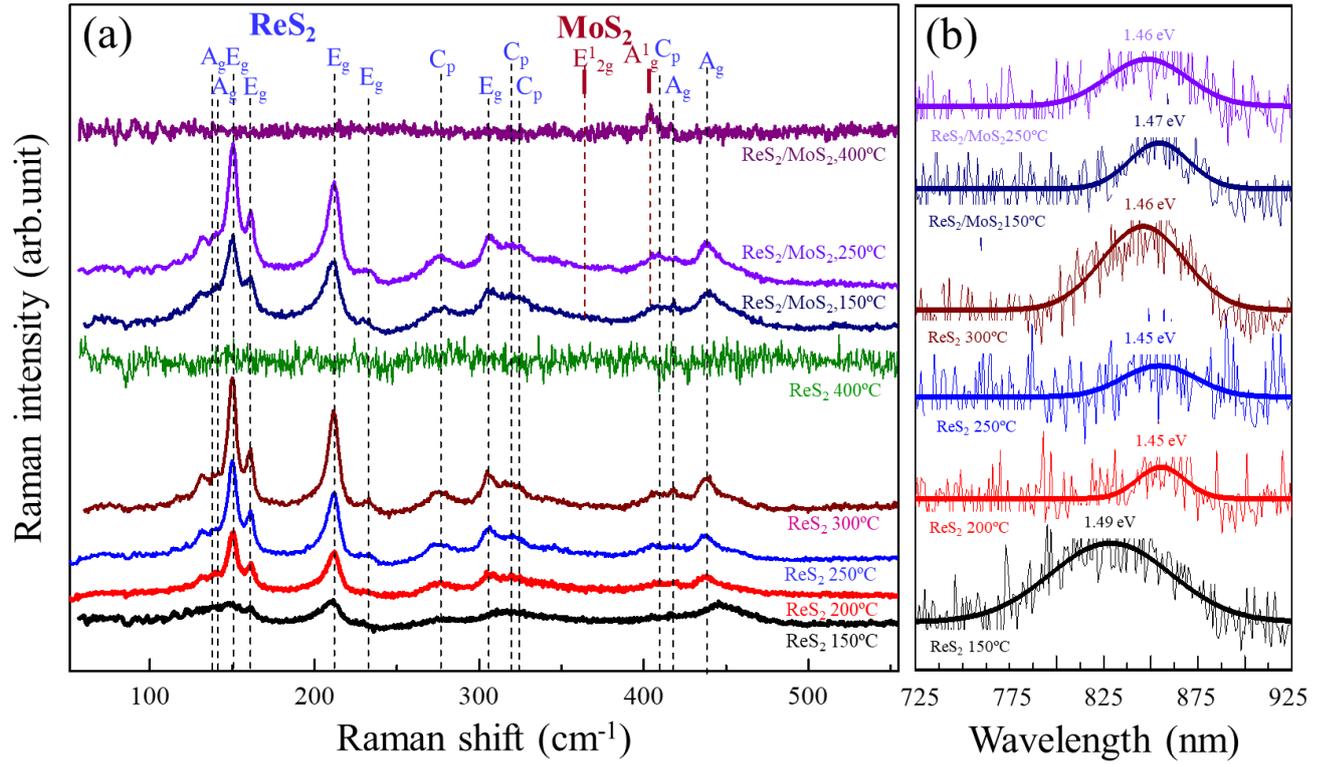

**Fig.6.** (a) Raman spectra of various ReS$_2$ thin films grown on sapphire substrate and MoS$_2$ template at different temperatures. No peaks corresponding to ReS$_2$ are observed for the films grown at 400°C. (b) CL spectra from various ReS$_2$ films showing emission peak around 1.46 eV.

# Growth of epitaxial ReS$_2$ thin film by pulsed laser deposition

B. Vishal,[1,*] H. Sharona,[1] U. Bhat,[1] A. Paul,[1] M. B. Sreedhara,[1] V. Rajaji,[1,3] S. C. Sarma,[2,3] C. Narayana,[1,3] S. C. Peter,[2,3] and R. Datta[1,*]

[1]International Centre for Materials Science, Chemistry and Physics of Materials Unit, Jawaharlal Nehru Centre for Advanced Scientific Research, Bangalore 560064, India.

[2]New Chemistry Unit, Jawaharlal Nehru Centre for Advanced Scientific Research, Bangalore 560064, India.

[3] School of Advanced Materials, Jawaharlal Nehru Centre for Advanced Scientific Research, Bangalore 560064, India.

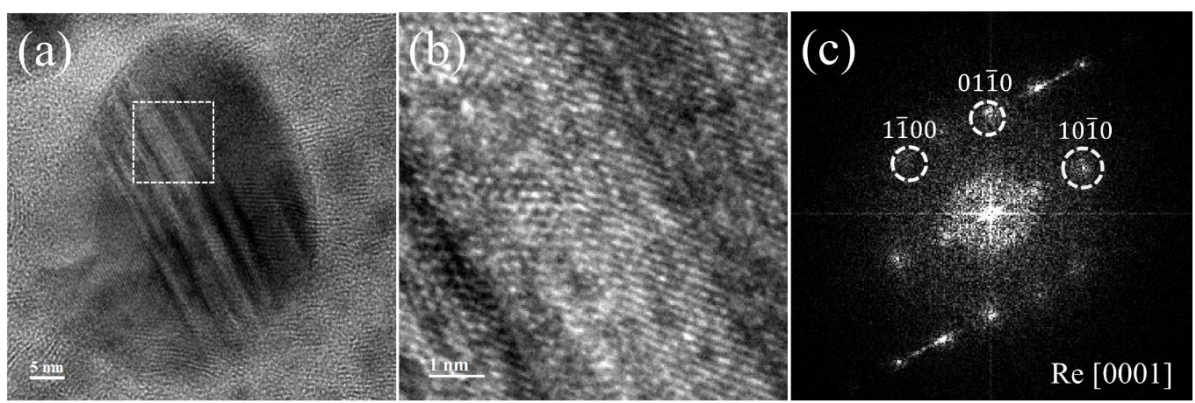

**Figure S1**. (a) Low maginification TEM image of Re particle embedded in film grown at 150°C, (b) zoomed in image of white square box of (a), (c) corresponding FFT pattern confirming hexagonal Re metal.

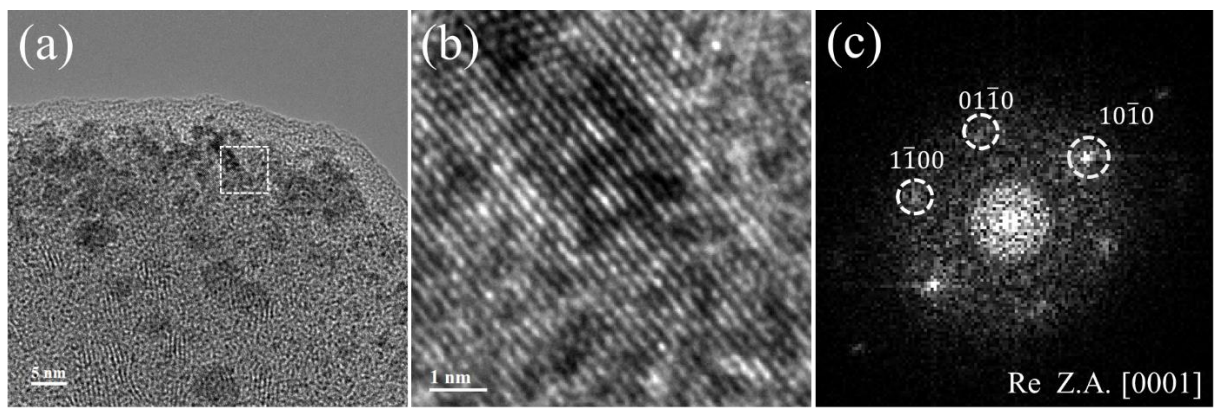

**Figure S2.**(a) Hexagonal Re Metals (~5 nm) at surface of T = 150°C Thin film surface. (b) zoomed in image of white square box of (a), (c) corresponding FFT pattern confirming presence of hexagonal Re metal.

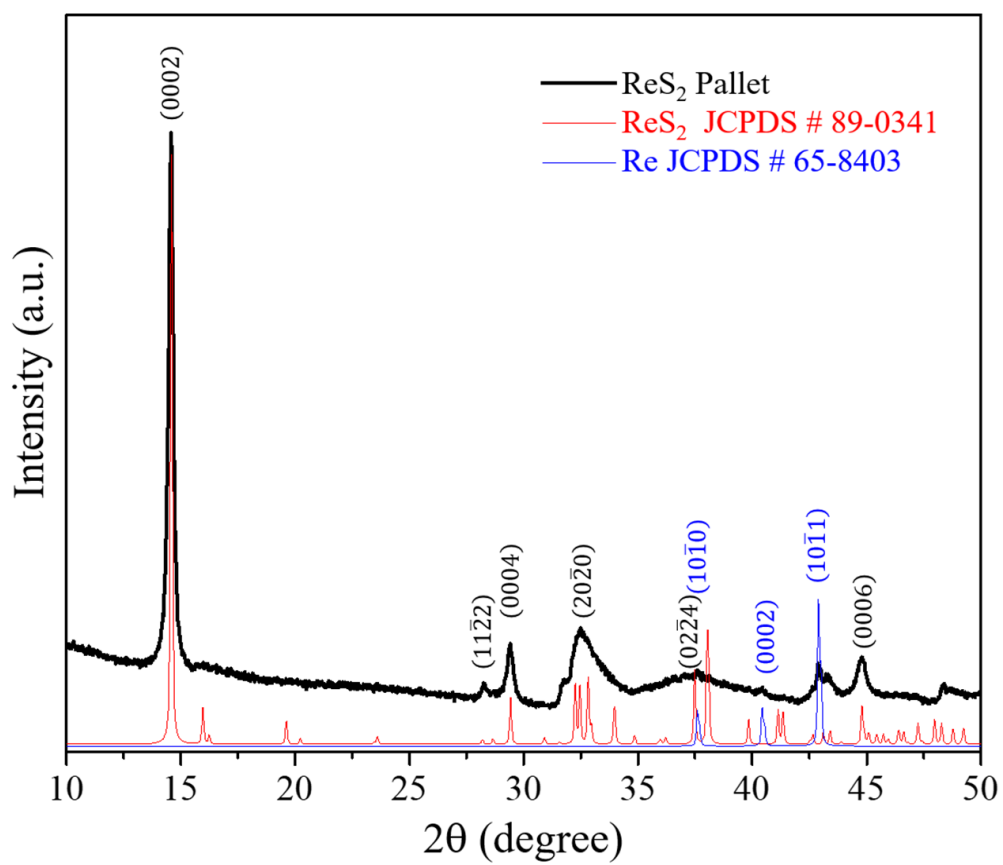

**Figure S3**. XRD of ReS2 pellet reveals minor presence of Re in the pallet.

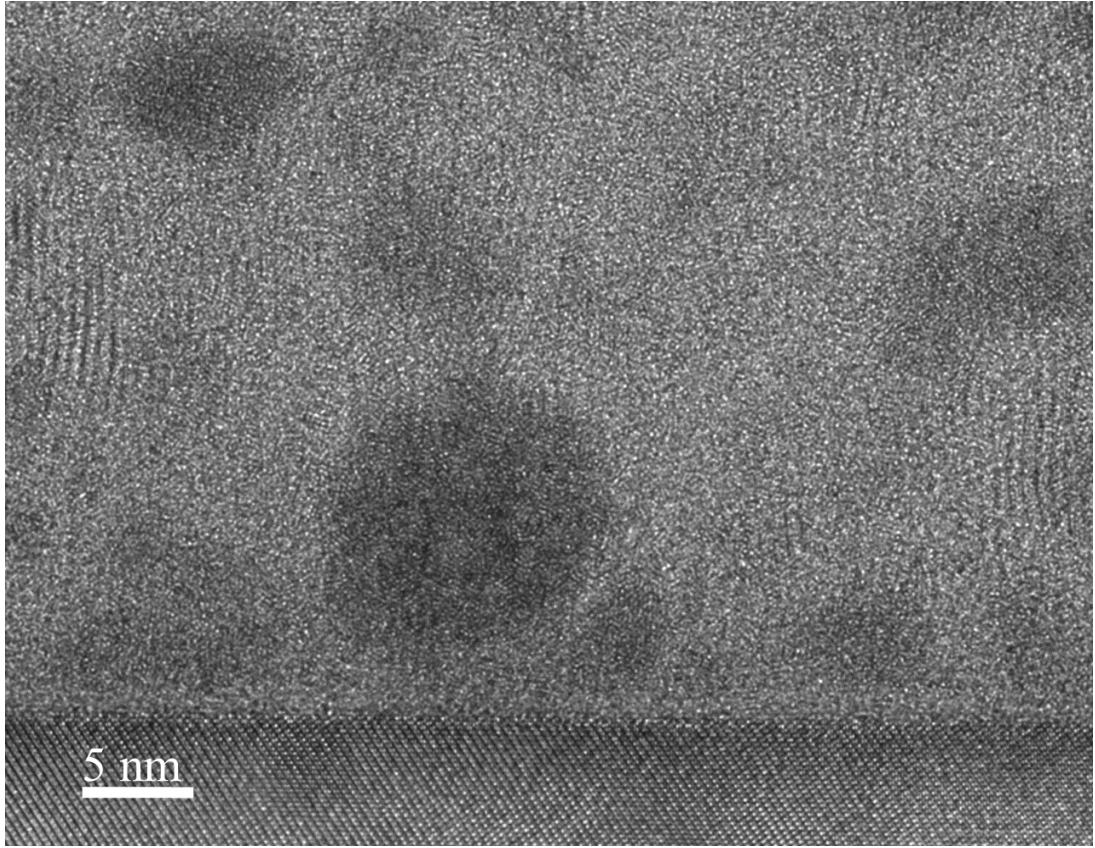

**Figure S4.** In some places of ReS$_2$ film grown at 150 °C films are found to be amorphous coexisting with the crystalline regions may be due to lower deposition temperature.

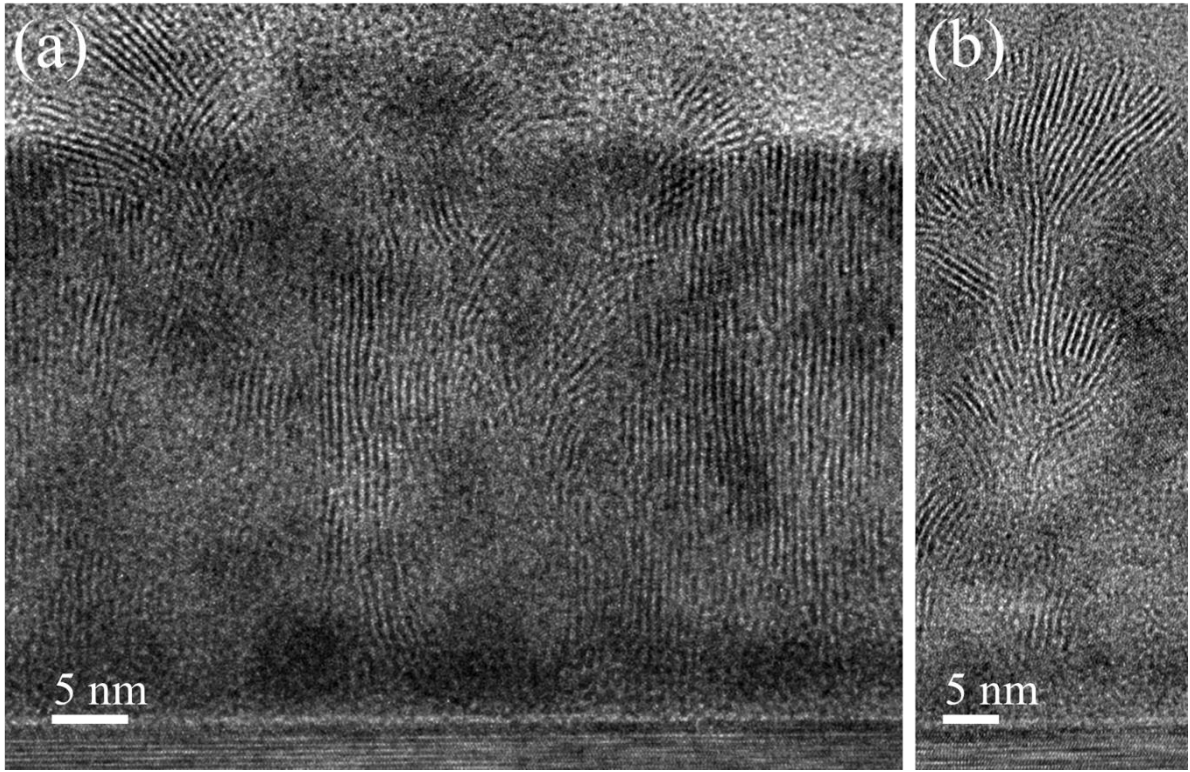

**Figure S5**. (a) & (b) From place to place the *c* planes start branching away from the vertical direction and creating a dome like structure at the surface for the film grown at 200 °C.

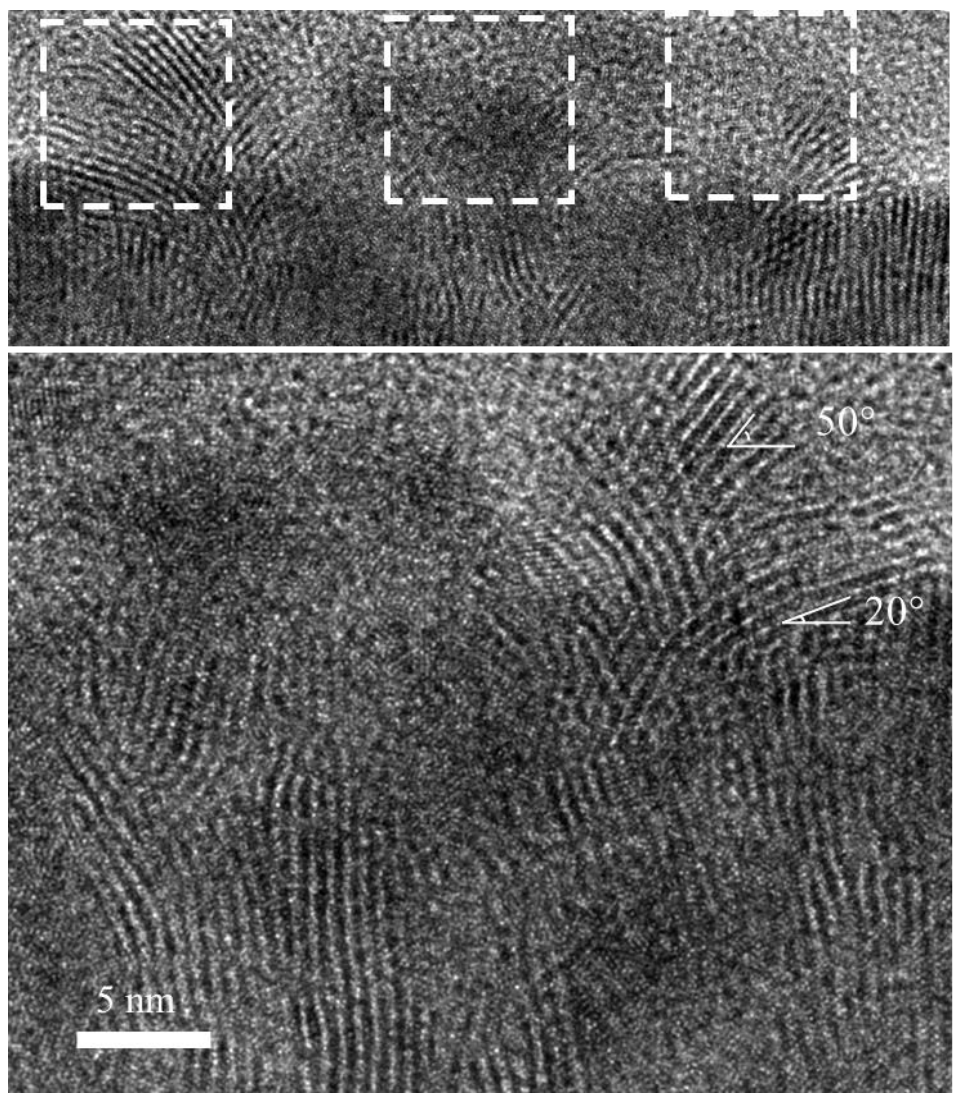

**Figure S6**. Branching in ReS$_2$ in several direction which reflect as different peaks in XRD pattern shown in figure 5.

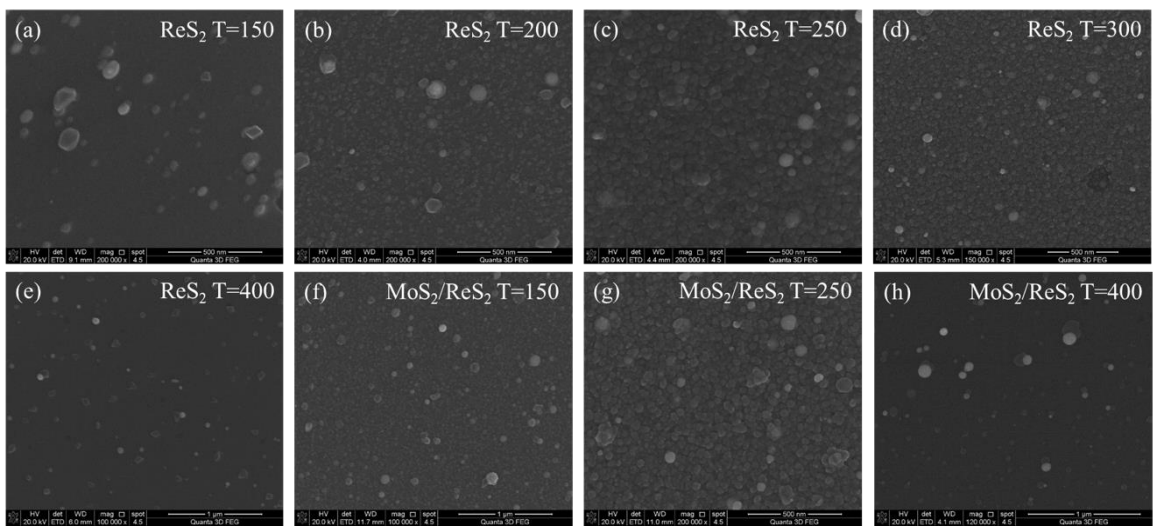

**Figure S7**. FESEM image show large area coverage of ReS$_2$ thin film on top of c-plan sapphire as well as MoS$_2$ template.